\documentclass[12pt]{article}

\begin{document}

\begin{titlepage}
\title{ On the Motion of Matter in Spacetime}

\author{T. Dereli,  R. W. Tucker\\
Department of Physics, Lancaster University,\\
Lancaster LA1 4YB, UK \\ {\small t.dereli@lancaster.ac.uk}\\
{\small r.tucker@lancaster.ac.uk}}
%\date{ 19  April  2001}

\maketitle

%\bigskip

\begin{abstract}
\noindent We argue that the geodesic hypothesis based on
auto-parallels of the Levi-Civita connection may need refinement
in  theories of gravity with additional scalar fields. This
argument is illustrated with a re-formulation of the Brans-Dicke
theory in terms of a spacetime connection with torsion determined
dynamically in terms of the gradient of the Brans-Dicke scalar
field.  The perihelion shift in the orbit of Mercury is calculated
on the alternative hypothesis that its worldline is an
auto-parallel of such a connection. If  scalar fields couple
significantly to matter and spinless test particles move on such
worldlines, current time keeping methods based on the conventional
geodesic hypothesis may need refinement.
\end{abstract}

\end{titlepage}

%\bigskip\bigskip
%\pacs{PACS no.: 04.20.Cv; 04.25.-g; 04.80.Cc; 96.30.Dz}

\bigskip

\def\Box{\triangle}
%\twocolumn
%\narrowtext
\def\pmb#1{\hbox{\bf #1}}
\def\bfr{{\pmb{r}}}
\def\bbfR{{\pmb{R}}}

\def\dotbfr{\dot{\bfr}}
\def\dotbbfR{\dot{\bfR}}
\def\dotbfR{\dot{\bfR}(\sigma,\eta)}

%---------------------------- differential geometry
\def\^{\wedge}
\def\pd{\partial} % partial derivative

\def\Lie{\hbox{\it \char'44}\!}

\def\bfT{{\bf {T}}}
\def\bfg{{\bf {g}}}

\def\bfd{\pmb{d}(\sigma,\eta)}
\def\bbfd{\pmb{d}}
\def\del{{ \nabla}}   % full covariant derivative

\def\LCdel{\hat \nabla}

\def\cs{ \left \{ {}^{\mu}_{\nu \lambda} \right \} }
%\def\csymbol#1#2#3{ \left \{ {}^#1_{#2\,#3} \right \} }

%\pagebreak
%\setcounter{page}{1}}

\def\GG{{\cal G}}
\def\TT{{\cal T}}
\def\delgzero{\nabla^{(g,0)}}
\def\delgT{\nabla^{(g,T)}}
\def\delgtzero{\nabla^{({\tilde{g}},0)}}

Einstein's formulation of interacting classical fields in
spacetime with a preferred geometry and his identification of
gravitation as a consequence of such a geometry is arguably one of
the most elegant mathematical descriptions of natural phenomena
ever produced. It has had a profound effect on our understanding
of local and global properties of the Universe and it remains an
inspiration in the quest for a unification of all the basic
interactions. It remains the paradigm for all successful
cosmological models and the benchmark for all tentative theories
of quantum gravity. The appropriate mathematical structure for the
formulation of classical gravitation is the  differential geometry
of a 4-dimensional manifold with a Lorentzian spacetime structure
and a connection on the bundle of linear spacetime frames. A
connection offers a means to compare  tensor fields  at different
events in spacetime  and generalises the familiar calculus of
derivatives. Once such a framework is in place, the physical
underpinnings of Einstein's theory of gravitation are surprisingly
simple to state. The theory takes as a fundamental tensor field a
symmetric  metric tensor $g$ and identifies gravitation with the
spacetime curvature associated with a particular connection
$\delgzero$ compatible with this metric. Such a geometry coexists
with {\it matter fields} according to the Einstein field
equations:
\begin{equation}
\GG(g,\delgzero)=\TT({\hbox{{\it matter fields}}},g).
\end{equation}
Einstein chose a connection that was determined uniquely by the
metric and so the Einstein tensor $\GG$ here also depends solely
on the metric. By contrast the influence of all other
``non-gravitational" fields (conveniently referred to as {\it
matter}) on gravity is determined by the structure of the
``stress-energy" tensor $\TT$. Not long after these equations were
formulated, Cartan\cite{cartan}, Weyl\cite{weyl},
Schr\"odinger\cite{schrodinger}  and others noted that more
general connections could be considered and that they offer a
means of describing geometrically the interaction of other fields
with gravitation. One such generalisation involves a connection
with a property called {\it torsion} by Cartan. Theories
formulated in terms of a  metric and a connection with torsion are
sometimes called Einstein-Cartan theories to distinguish them from
Einstein's original formulation in terms of a geometry with zero
torsion \cite{trautman}. Although it is always possible to rewrite
an Einstein-Cartan formulation in terms of Einstein's torsion free
choice of connection, the former sometimes offers definite
technical advantages when dealing with spinor fields and the use
of a general connection as a dynamical variable along with the
metric in variational formulations seems entirely appropriate.

In Einstein's theory  the effect of spacetime curvature on {\it
matter particles} is recognised as a {\sl gravitational force} and
it is commonly assumed that idealised massive spinless test {\it
particles} have spacetime histories that coincide with time-like
geodesics associated with the spacetime metric. Such assumptions
embody such notions as the ``equivalence principle'', the
``equality of inertial and gravitational mass" and the tenets of
``special relativity".  In terms of Einstein's torsion-free
connection compatible with the spacetime metric such particle
histories have self-parallel 4-velocities (tangent vectors) and
may be termed Levi-Civita auto-parallels. Assuming that our Sun
generates an exterior Schwarzshild metric in the vicinity of the
planet Mercury, one may use such hypotheses to calculate the
perihelion shift per revolution of its orbit and compare directly
with observation. Despite competing perturbations, this prediction
is regarded as one of the classical tests of any theory of
gravitation \cite{will} \cite{will2}. Further evidence for these
hypotheses comes from observations of the pulse rate of the
binary pulsar PSR 1913+16 that appears to be speeding up due to
gravitational radiation \cite{pulsar}.

Einstein, Infeld and Hoffmann \cite{einstein}
 made valiant attempts
to prove the {\it geodesic hypothesis} for test particles from a
field theory approach but their conclusions were not entirely
convincing. However, due to its naturalness, this hypothesis for
spinless test particles has almost become elevated to one of the
natural laws of physics \cite{weinberg}  and in
(pseudo-)~Riemannian spacetimes (where the geometry has zero
torsion) it arises convincingly as the lowest approximation to a
multipole expansion of matter distributions in tidal interaction
with gravity \cite{papapetrou}, \cite{dixon}.

However, there are compelling suggestions from astrophysical
observations that gravitational dynamics may require the inclusion
of hitherto undetected  fields of either gravitational or matter
origin. On the theoretical side modern ``low energy effective
string theories" are replete with scalar fields and most unified
theories of the strong and electroweak interactions predict such
fields with astrophysical implications \cite{brans}.

In 1961 Brans and Dicke \cite{BD}  suggested a modification of
Einsteinian gravitation by introducing a single additional scalar
field with particular gravitational couplings to matter via the
spacetime metric. They gave cogent arguments suggesting that the
experimental detection of such a new  scalar component to
gravitation might have been overlooked in traditional tests of
gravitational theories. Their theory is arguably the simplest
modification to Einstein's original description and in this note
we suggest that efforts to detect the effects of  scalar
gravitational interactions  with test particles of matter may have
overlooked a possibility that has new experimentally detectable
implications.

Brans and Dicke \cite{BD}  originally assumed the motion of a test
particle to be a Levi-Civita auto-parallel associated with the
metric derived from the Brans-Dicke field equations (even though
the scalar field could vary in spacetime). Later Dirac
\cite{Dirac} showed that (in a Weyl invariant generalisation) it
was more natural to generate the motion of a test particle from a
Weyl invariant action principle and that such a motion in general
differed from a Brans-Dicke Levi-Civita auto-parallel. Although
Dirac was concerned with the identification of electromagnetism
with aspects of Weyl geometry, even for neutral test particles it
turns out that test particles would follow auto-parallels of a
connection with torsion. In Ref. \cite{DT} we have shown that the
Brans and Dicke theory can be reformulated as a field theory on a
spacetime with dynamic torsion  $T$ determined by the gradient of
the Brans-Dicke scalar field $\Phi$:
\begin{equation}
 T = e^a \otimes \frac{d\Phi}{2\Phi}\otimes X_a  - \frac{d\Phi}{2\Phi}
\otimes {e^a}\otimes X_a
\end{equation}
in any coframe $\{e^a\}$ with dual $\{X_a\}$.
 Of course no new physics {\it of
the fields} can arise from such a reformulation. However, the
behaviour of spinless {\it particles} in such a geometry with
torsion depends a priori on the choice made from two possible
alternatives. One may assert that their histories are {\sl either}
geodesics associated with auto-parallels of the Levi-Civita
connection {\sl or}  the auto-parallels of the non-Riemannian
connection with torsion. Since one may find a spherically
symmetric, static vacuum solution to the Brans-Dicke theory (in
either formulation), it is possible to compare these alternatives
for the history of a mass in orbit about a spherically symmetric
source of scalar-tensor gravity by regarding it as a spinless test
particle as in General Relativity.

The  passage from a dynamic classical theory of fields to a theory
involving fields and particles in general depends on a number of
assumptions that are tantamount to a prescription of a model for
such  particles and their couplings to other fields. One such
approach can be based on the equations for the density and
velocity of a pressure-less fluid coupled to gravity. A particle
is then modelled as a bundle of integral curves of such a velocity
field and its motion follows as a consistency condition (using
Bianchi identities) for the postulated field equations for gravity
and the fluid.

In order to motivate an alternative to the conventional geodesic
hypothesis for spinless test particles based on the  Levi-Civita
connection, it is necessary to discriminate carefully between
alternative formulations of the underlying field descriptions of
such a fluid and appreciate the importance of establishing
fiducial coordinates and clocks for the comparison of experimental
data.
 Any proposed classical field theory of gravitation should determine a
spacetime metric consistent with a weak field Newtonian limit. All
experimental measurements of large scale gravitational phenomena
should be based on the use of clocks that measure (proper) time
with respect to such a spacetime metric. The behaviour of matter
is determined by the interaction of matter {\it fields } with
gravity. Massive test {\it particles} are idealisations that
neglect the influence of the particle on the geometry of
spacetime. Their collisionless motion can be determined either
from an action principle or an equation of motion for their
worldlines. Whether such actions or equations of motion can be
derived from the gravitational field equations depends on the
regularisation procedure used to model test particles and the
choice of stress tensor for distributed matter.

In general, gravitational fields can be expressed in terms of
different geometries involving connections that need  be neither
torsion-free nor compatible with the metric. The choice of
geometry is one of expediency in describing gravitational
interactions. In many cases gravitational interactions may be
conveniently encoded into non-Riemannian connections. In
mathematical terms, the motion of a particle is described by a
{\it parametrised curve} and a physical clock is used to fix the
parametrisation with respect to a choice of spacetime metric $g$.
Thus the 4-velocity $V_g$ of the particle obeys $g(V_g,V_g) =
-c^2.$

One may express the interaction of a test particle with
Brans-Dicke gravitation (involving a spacetime metric $g$ and a
scalar field $\Phi$) in terms of its equation of motion. In
succinct notation our test particle equation of motion \cite{DT2}
is given as
\begin{equation}
\nabla^{(g,T)}_{V_g}V_g = 0
\end{equation}
where $\nabla^{(g,T)}$ is a metric compatible connection
associated with the metric $g$ and the torsion $T$ above, and
$V_g$ is the velocity 4-vector of the particle  normalised with
respect to $g$. This normalisation implies that time measurements
are referred to a clock measuring proper time according to $g$.

In order to evaluate the motion of the test particle in a static
background geometry one must specify a static solution for $g$ and
$\Phi$. In general a family of metrics is available and one needs
to fix parameters in this family. If $g$ is asymptotically flat
one may take a weak field limit to identify a Newtonian potential
and Newtonian source mass via Poisson's equation for the
gravitational field. One may then require that the postulated
equation of motion for the particle yields Newton's law of motion
in the non-relativistic limit and furthermore maintain the
equivalence of gravitational and inertial mass. Thus a particular
system of coordinates is singled out to effect such
identifications and boundary conditions on static solutions. For
static spherically symmetric solutions, it is customary to choose
coordinates in which the spacetime metric takes the form
\begin{eqnarray}
g&=&-A(r) c^2 dt \otimes dt + B(r)^{-1} dr \otimes dr \nonumber \\
& & + r^2 ( d\theta \otimes d\theta + \sin^{2}\theta d\phi \otimes
d\phi) \label{eqn1}
\end{eqnarray}
where $A(r)$ and $B(r)^{-1}$ tend to $1$ for large $r$. It should
be stressed that the identification of the tensor $g$ with the
spacetime metric (determining the Newtonian potential) is pivotal
in this argument.

It is a mathematical fact that any theory written in terms of a
geometry with non-trivial $(g,T)$ and $\Phi$  can be reformulated
in terms of a geometry with either $(g,0)$ or $(\Phi g, 0)$.
Moreover it is also true that
\begin{equation}
\nabla^{(g,T)} = \nabla^{(\tilde{g},0)} - \frac{d\Phi}{2\Phi}
\label{eqn2}
\end{equation}
where $\tilde{g} = \Phi g$ with the choice of $T$ above and both
connections metric compatible ($\nabla^{(g,T)}g = 0 ,
\nabla^{(\tilde{g},0)}\tilde{g} = 0)$. This implies that, if
\begin{equation}
\nabla^{(g,T)}_{V_g}V_g = 0  \quad , \quad g(V_g,V_g) = -c^2 ,
\label{eqn3}
\end{equation}
then
\begin{equation}
\nabla^{(\tilde{g},0)}_{V_{\tilde{g}}}V_{\tilde{g}} = 0  \quad ,
\quad \tilde{g}(V_{\tilde{g}}, V_{\tilde{g}}) = -c^2 .
\label{eqn4}
\end{equation}
Thus one may map particle motions from one geometry to another.
However, the parametrised auto-parallels in different geometries
are strictly different. According to the above equations,
auto-parallels of $\nabla^{(g,T)}$ are parametrised with proper
time according to $g$ while the geodesics of
$\nabla^{(\tilde{g},0)}$ are parametrised with proper time
according to $\tilde{g}$. Since the perihelion rate of a test
particle depends on the proper time used to measure it, the choice
of metric is physically relevant. Furthermore by contrast to
(\ref{eqn1}) one would choose a different coordinate system to
express
\begin{eqnarray}
\tilde{g}&=&-\tilde{A}(\tilde{r}) c^2 dt \otimes dt +
\tilde{B}(\tilde{r})^{-1} d\tilde{r} \otimes d\tilde{r} \nonumber \\
& &  + {\tilde{r}}^2 ( d\theta \otimes d\theta + \sin^{2}\theta d\phi
\otimes d\phi) \label{eqn5}
\end{eqnarray}
in order to fix the parameters in $\tilde{g}$. Since $\tilde{g} =
\Phi g $ then $\tilde{r} \neq r$. It is interesting to note that
to first post-Newtonian order, the geodesics of $\tilde{g}$
measured by proper time with respect to $\tilde{g}$ are
indistinguishable from those based on the Schwarzschild metric.
However, one readily checks that this is not true for our equation
of motion (\ref{eqn3}) expressed in terms of the metric $g$ in
equation (\ref{eqn1}) above. Thus in order to meaningfully compare
different proper time parametrised orbits, both the choice of
coordinates and spacetime metric should be the same.

One may derive the equation of motion (\ref{eqn3}) in the context
of pressure-less fluid flow (approximating the flow of particles
of matter)  in a number ways. The natural choice of matter
stress-energy tensor is $\rho \,V_G \otimes V_G$ for some function
$\rho$ and velocity field $V_G$ normalised with respect to some
metric $G$. In their original formulation with a metric $g$, Brans
and Dicke coupled such matter so that the Bianchi identities
implied that the vector $V_g$ satisfied
\begin{equation}
{\nabla}^{(g,0)}_{V_g} V_{g} = 0 \label{eqn9}
\end{equation}
and conservation of the current $\rho V_g$ with $ g(V_g,V_g) =
-c^2.$ In our reformulation of their theory in terms of a
spacetime geometry $(g,T)$ with the torsion given above \cite{DT},
their field equations are
\begin{eqnarray}
\Phi\,\GG(g,\delgT) &=& \frac{\gamma}{\Phi}\,\TT(\Phi,g) +
\rho\,V_g\otimes  V_g ,   \nonumber  \\
\frac{\gamma}{2}\, \Box_g\Phi &=& -\rho
\end{eqnarray}
 where $\gamma = 4\omega +6$ with Brans-Dicke parameter
$\omega$. In this equation $\TT(\Phi,g)$ is proportional to the
canonical stress tensor for a ``massless Klein-Gordon" scalar
field $\Phi$.
 However, one can also
rewrite the vacuum Brans-Dicke equations in terms of the spacetime
geometry $(\tilde{g} = \Phi g, 0)$:
\begin{eqnarray}
\GG(\tilde{g},\delgtzero) &=& {\gamma}\,\TT(\ln \Phi,\tilde{g}) ,
\nonumber  \\
{\gamma}\, \Box_{\tilde{g}}\ln \Phi &=& 0
\end{eqnarray}
where  quantities with tilde refer to the geometry
$(\tilde{g},0)$.

An alternative way to couple a pressure-less fluid that reduces to
the above in the absence of matter is to postulate the matter
coupling with
\begin{eqnarray}
\GG(\tilde{g},\delgtzero) &=& {\gamma}\,\TT(\ln \Phi,\tilde{g}) +
\hat\rho\,V_{\tilde{g}}\otimes  V_{\tilde{g}} , \nonumber  \\
{\gamma}\, \Box_{\tilde{g}}\ln\Phi &=& 0 \label{eqn12}
\end{eqnarray}
where $\tilde{g}(V_{\tilde{g}},V_{\tilde{g}}) =-c^2.$ In this case
the Bianchi identities imply  conservation of the current
$\hat\rho V_{\tilde{g}}$  and the equation of motion
\begin{equation}
{\nabla}^{(\tilde{g},0)}_{V_{\tilde{g}}}V_{\tilde{g}} = 0
\label{eqn13}
\end{equation}
 which
 we have shown above  is equivalent to the auto-parallel equation of
 motion (\ref{eqn3})
 and  distinct from that following from the matter coupling
assumed by Brans-Dicke above. Indeed,  in terms of the spacetime
geometry $(g,T)$  the field
 equations (\ref{eqn12}) may be rewritten as
\begin{eqnarray}
\Phi\,\GG(g,\delgT) &=& \frac{\gamma}{\Phi}\,\TT(\Phi,g) +
\Phi^{{1}/{2}}\hat\rho\,V_g\otimes  V_g , \nonumber  \\
{\gamma}\, \Box_g\Phi &=& 0
\end{eqnarray}
 where $g(V_g,V_g) = -c^2.$
 As expected the Bianchi identities reproduce
the same matter equation of motion but expressed in the form
\begin{equation}
{\nabla}^{(g,T)}_{V_g} V_{g} = 0 . \label{eqn15}
\end{equation}

\def\eh0{\hat\epsilon}
\def\rh0{\hat r_0}

The solutions of this equation are {\it auto-parallels} of
${\nabla}^{(g,T)} .$ They coincide with the geodesics of
(\ref{eqn13}) (auto-parallels of ${\nabla}^{(\tilde{g},0)}$).
Comparing this with the equation for the standard Levi-Civita
geodesics (\ref{eqn9}) (auto-parallels of ${\nabla}^{(g,0)}$), we
note that any experiment that depends only on the conformal metric
structure of spacetime cannot distinguish solutions of
(\ref{eqn15}) from (\ref{eqn9}). The bending of starlight by the
gravitational fields falls into this category. However, the motion
of {\it massive} test particles offers a means to test the new
effects of scalar field interactions according to (\ref{eqn15}).
To illustrate the effects of this equation of motion we have
recomputed \cite{DT2} the classical shift in the perihelion rate
of Mercury's orbit about the Sun in terms of a static spherically
symmetric solution of the vacuum Brans-Dicke field equations
\cite{BD}.
 Taking into account that the speed of Mercury is
non-relativistic and that its Newtonian orbit is much larger than
the Schwarzschild radius $r_s=2GM/c^2$ of the Sun, one finds
 the perihelion shift per revolution
$\Delta$ of the orbit:
\begin{eqnarray}
\Delta=\frac{3\omega+5}{3\omega+6}\,\,\delta_\omega \label{ans1}
\end{eqnarray}
where $\delta_\omega=3\lambda_\omega\pi$ and
$\lambda_\omega={r_s}/{\rh0}$. Using the Kepler period $\hat
T=2\pi({\rh0^3}/{(1-\eh0^2)GM})^{1/2}$, the shift $\delta_\omega$
may be expressed in terms of $\hat T$ and $\eh0$. This may be
compared with the result based on the assumption that Mercury's
orbit follows from a geodesic of the torsion-free Levi-Civita
connection. In this case one finds
\begin{eqnarray}
\Delta=\frac{3\omega+4}{3\omega+6}\,\,\delta\label{ans2}
\end{eqnarray}
where $\delta=3\lambda\pi$ (with $\lambda={r_s}/{{r_0}}$)
 is the  perihelion shift
per revolution of the orbit based on the Schwarzschild solution
for the metric in General Relativity \cite{will}.

In conclusion we maintain that in a  scalar-tensor theory based on
the Brans-Dicke field equations for a static spherically symmetric
spacetime metric $g$, the perihelion shift of a test particle
calculated in the standard coordinates given above according to
(\ref{eqn1})  and (\ref{eqn15}) differs from the original
hypothesis of Brans and Dicke (based on $\nabla^{(g,0)}_{V_g}V_g =
0  , \, g(V_g,V_g) =-c^2$) and also from that based on
(\ref{eqn4})  in coordinates with radius $\tilde{r}$ adapted to
the conformally related
 metric $\tilde{g}$.

Given the prominent role played by the motion of test particles in
many astrophysical phenomena and the possibility of new
gravitational interactions mediated by scalar fields we feel that,
with the enhanced technology now available to modern space
science, the possible relevance of  scalar field induced spacetime
torsion should not be ignored. Should the departure  of spinless
particles  from Levi-Civita geodesic motion be detected in some
purely gravitational environment it would indicate that matter has
additional ``gravitational charge" in addition to its mass and to
its  electromagnetic, weak and strong couplings to other fields in
nature.

\bigskip

\noindent {\bf Acknowledgement}

We are grateful to  BAE-Systems (Warton), PPARC and the Leverhulme
Trust for financial support. We also thank Dr. Robert Low for
useful comments.

%\vskip 0.6cm \vskip 0.6cm \vfil\eject

%%%%%%%%%%%%%%%%%%%%%%%%%%%%%%%%%%%%%%%%%%%%%%%%%%%%%%%%

\vfil\eject
\end{document}